\pdfoutput=1
\documentclass[10pt,conference]{IEEEtran}

\usepackage[utf8]{inputenc}
\usepackage{cite}
\usepackage{ifpdf}
\usepackage{subcaption}
\usepackage{graphicx}
\usepackage[braket, qm]{qcircuit}
\usepackage{xcolor}
\usepackage{soul}
\usepackage{tabularx}
\usepackage{array}
\usepackage{lipsum}
\usepackage{multicol}
\usepackage{gensymb}
\usepackage[hidelinks]{hyperref}

\usepackage{balance}

\newcommand{\jakarta}{\texttt{ibmq\char`_jakarta} }
\def\Scale{0.8}
\def\RowHeight{0.2}
\def\Width{1.0}

\begin{document}

\pagestyle{plain}

\title{Extending and Defending Attacks\\ on Reset Operations in Quantum Computers}

\author{

\IEEEauthorblockN{Jerry Tan}
\IEEEauthorblockA{
\textit{Yale University}\\
New Haven, CT, USA \\
jerry.tan@yale.edu
}

\and

\IEEEauthorblockN{Chuanqi Xu}
\IEEEauthorblockA{
\textit{Yale University}\\
New Haven, CT, USA \\
chuanqi.xu@yale.edu
}

\and

\IEEEauthorblockN{Theodoros Trochatos}
\IEEEauthorblockA{
\textit{Yale University}\\
New Haven, CT, USA \\
theodoros.trochatos@yale.edu
}

\and

\IEEEauthorblockN{Jakub Szefer}
\IEEEauthorblockA{
\textit{Yale University}\\
New Haven, CT, USA \\
jakub.szefer@yale.edu
}

}

\date{}
\maketitle

\IEEEpeerreviewmaketitle

\begin{abstract}
  The development of quantum computers has been advancing rapidly in recent years. As quantum computers become more widely accessible, potentially malicious users could try to execute their code on the machines to leak information from other users, to interfere with or manipulate the results of other users, or to reverse engineer the underlying quantum computer architecture and its intellectual property, for example. Among different security threats, previous work has demonstrated information leakage across the reset operations, and it then proposed a secure reset operation could be an enabling technology that allows the sharing of a quantum computer among different users, or among different quantum programs of the same user. This work first shows a set of new, extended reset operation attacks that could be more stealthy by hiding the intention of the attacker's circuit. This work shows various masking circuits and how attackers can retrieve information from the execution of a previous shot of a circuit, even if the masking circuit is used between the reset operation (of the victim, after the shot of the circuit is executed) and the measurement (of the attacker). Based on the uncovered new possible attacks, this work proposes a set of heuristic checks that could be applied at transpile time to check for the existence of malicious circuits that try to steal information via the attack on the reset operation. Unlike run-time protection or added secure reset gates, this work proposes a complimentary, compile-time security solution to the attacks on reset~operation.
\end{abstract}

\section{Introduction}
\label{sec_introduction}

Noisy Intermediate-Scale Quantum (NISQ) quantum computers are being rapidly developed, with machines over $400$ qubits available today~\cite{433qubit} and the industry projects $4000$-qubit or larger devices before the end of the decade~\cite{4000qubits}. Many different types of quantum computers exist, with superconducting qubit quantum computers being one of the types available today to researchers and the public through cloud-based services. The superconducting qubit machines are developed by numerous companies, such as IBM~\cite{ibmquantum}, Rigetti~\cite{rigetti}, or Quantum Circuits, Inc.~\cite{quantumcircuitsinc}.

Quantum computers of these sizes have the potential to fundamentally alter what types of algorithms can run on them, but require specialized facilities and equipment in order to make these quantum computers accessible to users. There is a growing interest in, and practical deployments of, cloud-based quantum computers, also called Quantum as a Service (QaaS) or Quantum Computing as a Service (QCaaS). Cloud-based services such as IBM Quantum, Amazon Bracket, and Microsoft Azure already provide access to quantum computers remotely for users. Following the past success of classical computer cloud-based services, we expect that cloud-based access for remote users to rent quantum computers to be a dominant use-case in the future.

In order to support sharing of a quantum computer among different users, there needs to be an efficient way to reset the qubits. Today, the main method to reset the qubit state is by letting qubits decohere, which allows qubits to naturally decay into their ground states. Even though letting qubits decohere erases all the qubit states, it takes a long time, i.e., 250 ns is required for quantum computers on IBM Quantum; it also makes the qubits unusable during that time. As an alternative, a number of companies, such as IBM, have proposed a reset gate or reset operation. The reset operation first measures the qubit state, which collapses it to $\ket 0$ or $\ket 1$ based on the state of the qubit. Next, if the qubit collapsed into $\ket 1$, an {\tt X} gate (similar to classical {\tt NOT} gate) is applied to set the qubit state to $\ket 0$ state, and the qubit is now fully reset.

Mi et al.~\cite{mi2022securing}, however, explored the existing (insecure) reset operations used in superconducting quantum computers such as from IBM Quantum and showed that they do not protect fully from information leakage since the reset operation is not perfect. Since the reset operation is conditional on measurement results, its outcomes are closely associated with the error characteristics of the measurement operation. As it was shown~\cite{mi2022securing}, an attacker measuring the qubit state post-reset can statistically recover some information about the qubit’s state prior to the reset, thus leaking information from the victim user who was using the same qubit prior to the attacker. The fundamental idea behind their attack circuit was for the attacker to perform a qubit measurement immediately when scheduled to execute. Such a malicious circuit, however, can very easily be detected since it only contains a measurement~gate.

Our work proposes a new, extended attack on reset operations. In particular, our work explores potential ways in which an attacker can add a masking circuit $C$ before the measurement to ``hide'' their attack. The main idea behind our design is that by using a masking circuit $C$ the attacker can make their circuit look like a benign circuit while still being able to recover information across the reset operation as before. In particular, we show that an attacker can use a large number of circuits to target a particular qubit for information leakage, as long as the attacker's circuit is composed of single qubit operations on the target qubit. The attacker can also hide their intention and attack by using two-qubit {\tt CX} gates, as long as the target qubit of the attack is the control qubit of the {\tt CX}~gates.

For single-qubit gates used in the masking $C$ circuit, the attacker may use simple identity circuits consisting of pairs of {\tt X} gates, or non-identity circuits consisting of as {\tt RX} and {\tt RZ} gates. For multi-qubit gates, an attacker can also hide an attack with {\tt CX} gates, as long as the target qubit is the control qubit of the {\tt CX} gate. We also show conditions under which the attack becomes more difficult, such as when qubits are targets of {\tt CX} gate. We confirm our expectation by running select QASM benchmark circuits, and showing that it is difficult for the attacker to leak the victim's state, due to the presence of multi-qubit gates or other non-identity gates, if the masking circuit $C$ is a full QASM benchmark, for example.

Based on our findings and possible new attacks, we present a new set of heuristics defenses that could be applied to check for existence of the new kind of the malicious circuits before code is executed.
Unlike run-time protection or added secure reset-gates, this work proposes a complimentary, compile time security solution to the attacks on reset~operation. Note, that previous work~\cite{mi2022securing} proposed a secure reset gate for use at run-time, while we propose a compile-time defense. Our solution meanwhile draws inspiration from different previous work~\cite{deshpande2023design} which proposed a quantum computing antivirus that aim to flag suspicious programs that inject malicious crosstalk and degrade the quality of program outcomes. Instead of focusing on crosstalk, we explore how to check circuits for malicious reset operation attacks. Instead of focusing on graph structure of the circuit, we provide a solution based on calculating the matrix representation of the circuit (where possible due to circuit size) as well as based on analyzing types of gates execution on each qubit within a circuit.

\subsection{Contributions}

The main contributions of this work are as follows:

\begin{itemize}
    \item Presentation of a new variant of attacks on reset operations, involving a masking circuit used by the attacker to try to hide their attack circuit.
    \item Evaluation of the efficacy of different masking circuits in the new attack variant.
    \item Description of a set of heuristics to detect existing and the new attacks on reset~operation.
    \item Demonstration of a tool and compile-time approach tool for detection of previous attacks and the new attack variant using the heuristics.
\end{itemize}

\section{Background}
\label{sec_background}

Qubits are the fundamental building blocks of quantum computers. They encode data in quantum states, which can exist as a superposition, and are able to represent a continuum of states in between the classical $0$ and $1.$ To observe the state of a qubit, the qubit state must be collapsed by a measurement operation, also known as a readout. The two possible measurement results are $0$ and $1$, corresponding to eigenstates $\ket{0}$ and $\ket{1}$.

\subsection{Bloch Sphere}

The Bloch sphere is a geometric representation of a two-level quantum system. It provides a way to visualize an arbitrary state of a qubit as a superposition of the two computational basis vectors, $\ket{0}$ and $\ket{1}$. The surface of the Bloch sphere can be parameterized by two angles used in the spherical coordinate system: $\theta$ with respect to the $z$-axis, and $\phi$ with respect to the $x$-axis. Given angles $\theta, \phi,$ we write the corresponding quantum state:

$$\ket{\psi}=\cos\left(\frac{\theta}{2}\right)\ket{0}+e^{i\phi}\sin\left(\frac{\theta}{2}\right)\ket{1},$$ 

\noindent where $0\leq \theta \leq \pi$ and $0\leq \phi<2\pi.$ Quantum circuits are mainly composed of gate operations, also simply called gates, which can be visualized as applying various rotations of the quantum state around the Bloch sphere.

\subsection{Basis Gates}

Quantum gates are used to manipulate quantum states. Reversible operations can be represented by unitary matrices, and quantum gates exist for various unitaries. For each quantum computer, some gates are supported as a native gates, also called basis gates by IBM, for example. Most NISQ quantum computers, including IBM machines, only support a few native gates: the single-qubit gates ({\tt I, RZ, X, SX}), and one two-qubit gate ({\tt CX}). Other gates need to be decomposed into these basis gates first before being run on the~machines.

Among single-qubit gates, {\tt I} is the identity gate, that performs no operation, but adds delay. The {\tt X} gate performs a rotation around the $z$ axis of the Bloch sphere by a fixed $\pi$ radians angle for the target qubit. It is also analogous to the classical {\tt NOT} gate, as it maps $\ket{0}$ to $\ket{1}$ and $\ket{1}$ to $\ket{0}$, thus ``flipping'' the qubit. The {\tt RZ} gate performs a rotation of $\phi$ radians around the $z$ axis in the Bloch sphere for the target qubit. The {\tt SX} gate rotates a qubit around the $x$-axis a fixed angle of $\pi/2$ radians, it effectively adds the rotation angle to $\theta$ in the Bloch sphere for the target qubit.

For two-qubit gates, the {\tt CX} gate is available. The {\tt CX} gate operates on two qubits: a control qubit and a target qubit. If the control qubit is in state $\ket{0},$ the {\tt CX} acts as identity. Otherwise, if the control qubit is in state $\ket{1},$ an {\tt X} gate is applied to the target qubit, flipping it. The {\tt CX} gate is sometimes called the {\tt CNOT} gate.

\subsection{RX Gates}

The {\tt RX$(\theta)$} gate performs a rotation of $\theta$ radians around the $x$-axis of the Bloch sphere. The {\tt RX} gate is not a native gate, but it can be decomposed into native basis gates {\tt RZ} and {\tt SX}~gates.

\subsection{Measurement Operation}

When a qubit is measured, the result is a classical bit of information, either $0$ or $1$. The measurement process collapses the original qubit state, projecting it typically onto the $z$-axis of the Bloch sphere. Measurement results of $0$ and $1$ correspond to state collapse into $\ket{0}$ and $\ket{1},$ respectively. Measurement is an example of a non-unitary operation, as it cannot be reversed. This state collapse is irreversible; after a measurement is made, the original information about the qubit of the state is lost. 

For a general qubit state $\ket{\psi}=\cos\left(\frac{\theta}{2}\right)\ket{0}+e^{i\phi}\sin\left(\frac{\theta}{2}\right)\ket{1},$ the collapse is probabilistic. The probability of a measurement is the square of the magnitude of the coefficient of the corresponding eigenstate. So we measure $0$ and $1$ with probabilities $\cos^2(\theta/2)$ and $\sin^2(\theta/2),$ respectively. For example, if $\theta$ is $\pi/2$, then probability of $0$ and $1$ being measured should be $50$\%.

\subsection{Reset Operation}

Another non-unitary operation is the reset operation. The reset operation consists of first making a measurement of a qubit onto a classical bit $c$. Then, an {\tt X} gate is conditionally applied to the qubit if classical bit $c$ measures $1$. In more detail, the measurement collapses the qubit to either the $\ket{1}$ or $\ket{0}$ state. In the former case, the classical bit reads $1$, and an {\tt X} gate is applied to, resulting in the $\ket{0}$ state. In the latter case, no {\tt X} gate is applied and the qubit remains in $\ket{0}.$

However, this design of the reset operation is susceptible to readout errors by the measurement operation. If a $\ket{1}$ is mistakenly read as $0$ or a $\ket{0}$ as a $1$, the reset operation incorrectly produces a final state of $\ket{1}.$ This error on the real machines leads to possible information leak to a malicious user on the same qubit \cite{mi2022securing}.

\subsection{Transpilation Process}

Transpilation is the process of transforming an input circuit for execution on specific hardware. It involves matching the circuit to the topology of a quantum device and decomposing the user's gates into native gates supported by the hardware. Similar to classical compilers, transpilers also optimize the programs for performance. Optimizations may involve rewriting non-linear flow logic, processing iterative sub-loops, conditional branches, and other complex behaviors.

\section{Extending Quantum Computer Reset Gate Attacks}
\label{sec_attack_model}

\begin{figure}[t]
    \centering
    \includegraphics[width=0.6\columnwidth]{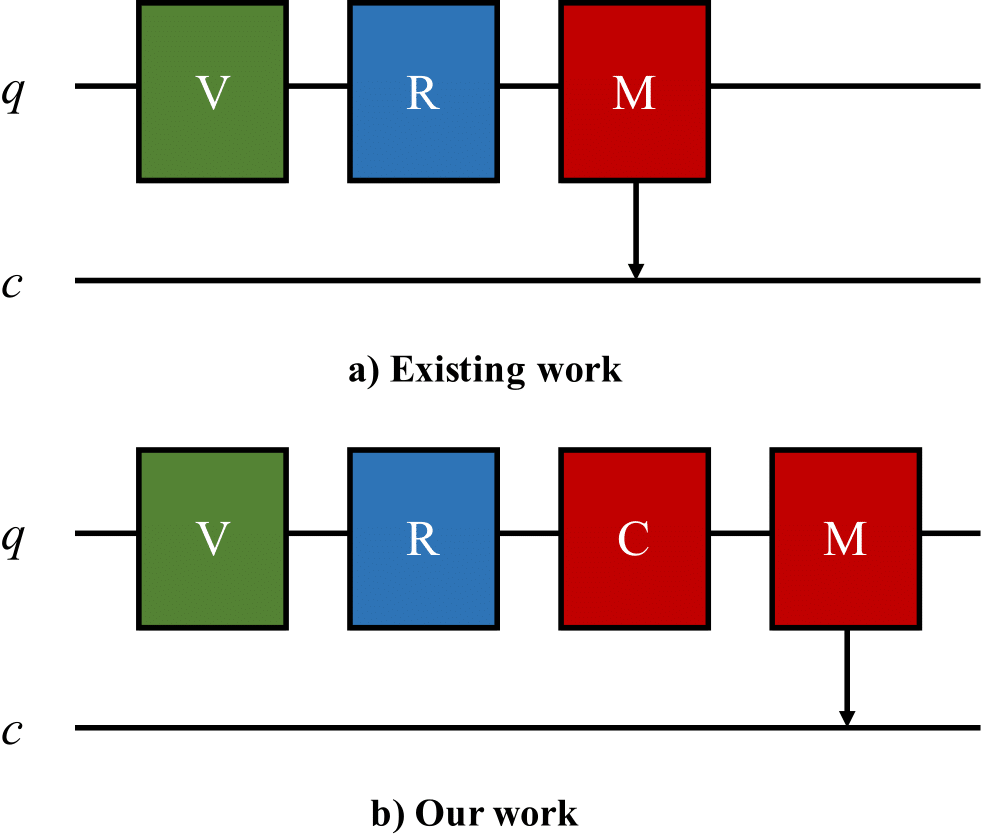}
    \caption{\small Attack model, $q$ represents target qubit and $c$ represents its corresponding classical register. $V$ is a shot of victim's circuit, $R$ is an inter-shot qubit reset mechanism, $C$ is a masking circuit used by attacker, and $M$ is the measurement operation used by attacker to try to guest the state of the $V$ before $R$.}
    \label{fig_attack_model}
\end{figure}

Previous work by Mi et al.~\cite{mi2022securing} has demonstrated information leak across the reset operation on IBM Quantum computers. A malicious attacker can use a circuit consisting of just a measurement gate on the same qubit as a victim to extract information about the amplitude of the $\ket{1}$ state, or the equivalent $\theta$ angle, of the victim state before reset. We assume that a strong attacker is able to run their program immediately after the victim, on the same qubits that the victim used. We also assume the qubits used by the victim are reset before the attacker can access them. Before the victim's reset, we assume the victim likely ends their computation with a measurement on all involved qubits. This collapses the victim qubit states to either $\ket{0}$ (where $\theta=0$) or $\ket{1}$ (where $\theta=\pi$). This scenario is most advantageous for the attacker since they only need to distinguish the two ends of the measured output frequency~distribution.

It has been shown in the prior work that even with multiple reset gates before the attack, information leak still occurs. The attacker model of the prior work is shown in Figure~\ref{fig_attack_model}a. In the figure, $V$ represents the victim circuit, which includes the victim's final measurement. $R$ represents one or more reset operations executed as a reset sequence between shots of circuit. $C$ represents the attacker's masking circuit, and $M$ represents the attacker's measurement.

However, a very simple defense mechanism can easily detect such an attack: scan for user circuits consisting of only one measurement gate, or more generally any circuit that begins with a measurement gate and flag these as suspicious.

This work shows that an attacker can bypass such simple defenses, and also make a more potent attack circuit, by adding a masking circuit $C$ before the measurement. By using a masking circuit $C$, the attacker can make their circuit look like a benign quantum circuit, but still be able to extract information across the reset operation as before. This work shows various masking circuits and how attackers can recover information even if the masking circuit $C$ is between the reset operation (of the victim) and the measurement (of the attacker). The high-level idea behind the extended quantum computer reset gate attacks is that the masking circuit $C$ represents unitary operations which can be reversed. With knowledge of the measurement and the masking circuit, the attacker can gain information about the state right before the masking circuit, which is related to the victim's state right before the reset. The attack model is shown in Figure~\ref{fig_attack_model}b.

\subsection{Attack Objective}

The first objective of this research work is to analyze the different types of masking circuits $C$ that an attacker could utilize. By using various masking circuits, the attacker can make their circuit look like a benign circuit, making detection of the attack harder, while at the same time still being able to carry out the reset gate attack where some information is learned about the state of the qubits prior to the~reset.

\subsection{Attacker Circuits}

This work explores and analyzes a variety of possible masking circuits $C$. Later we show which ones work well, and which ones do not.

\begin{itemize}
    \item Identity Circuits -- circuits consisting of an even number of single-qubit {\tt X} gates on each qubit, such that the total effective angle of rotation $\theta$ is $0$. Since effectively there is no rotation, the attacker's measurement should return the same values as it would be right after the reset operation. 
    \item {\tt RX} and {\tt RZ} Gate Circuits -- circuits consisting of single-qubit gates with effective $\theta$ ({\tt RX} gate) rotation and $\phi$ ({\tt RZ} gate) rotation. Because the rotation angle is known, the attacker can infer the qubit $1$-output probabilities as they would be right after the reset gate based on their measurement. As we demonstrate, certain rotation angles make the attack more difficult, while others still allow the attacker to make a meaningful measurement.
    \item {\tt CX} Gate Circuits -- circuits consisting of two-qubit {\tt CX} gates where there is entanglement between qubits. The control qubits of {\tt CX} gate experience delay (due to duration {\tt CX} gate) but otherwise can be leveraged by an attacker since they do not experience any rotations; meanwhile, the state of the target qubits of {\tt CX} gate depends both on the prior state and the control qubit, making attacker's use of that qubit more difficult.
    \item QASM Benchmarks -- circuits from the QASM benchmark suite \cite{li2022qasmbench} which are real quantum computing circuits. These include the 2- and 3-qubit Grover search circuits and the 4-qubit quantum random number generator (QRNG).
\end{itemize}

\subsection{Hiding Reset Operation Attack with Identity Circuits}

\begin{figure}[t]
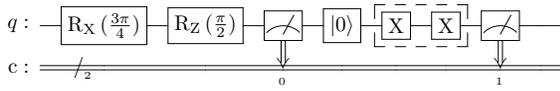

    \centering
    \include{figures/example_circuits/X_gates1}
    \caption{\small Example of two {\tt X} gate circuit used as a masking circuit; any even number of {\tt X} gates applied in sequence forms an identity circuit and can be evaluated for efficacy of the masking~circuit.}
    \label{fig_attack_identity}
\end{figure}

First, we experimented with using a series of {\tt X} gates as the attacker circuit, as shown in Figure~\ref{fig_attack_identity}. For a variety of input states, we ran experiments increasing the number of reset gates and the number of {\tt X} gate pairs, which we call the depth of the circuit. Since we use an even number of {\tt X} gates, the masking circuit is thus always equivalent to identity in this experiment group. As shown later in the Figures \ref{6_resets} and \ref{4x_snr}, information leak still occurs with {\tt X} gates added as a masking circuit. Based on the measured $1$-output frequency, the attacker can distinguish with high probability between victims initialized with $\theta = 0$ or $\theta = \pi$. 

An attacker may try more complex, non-identity circuits, or try to attack victims after a larger number of reset gates to avoid detection. We explain these next.

\subsection{Hiding Reset Operation Attack with {\tt RX} and {\tt RZ} Gate Circuits}

\begin{figure}[ht]
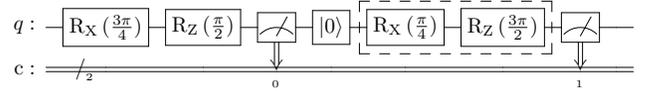

    \centering
    \include{figures/example_circuits/rx_rz_fixed_depth1}
    \caption{\small Example of masking circuit with {\tt RX} and {\tt RZ} gates, different number of {\tt RX} and {\tt RZ} gates and the angles can be evaluated for efficacy of the masking circuit.}
    \label{fig_attack_rx_rz_angle}
\end{figure}

Next, we considered {\tt RX} and {\tt RZ} rotation gates for the attacker to mask the attack. We ran two experimental groups. For the first set of attacks, we fixed the attack circuit depth at $1$ {\tt RX} and $1$ {\tt RZ} gate, and we varied the rotation angles. An example is shown in Figure~\ref{fig_attack_rx_rz_angle}.

\begin{figure}[t]
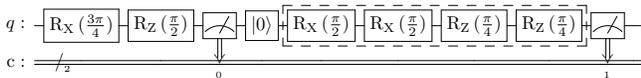

    \centering
    \include{figures/example_circuits/rx_rz_fixed_angle1}
    \caption{\small Example of different masking circuit with {\tt RX} and {\tt RZ} gates where the total rotation angles are fixed.}
    \label{fig_attack_rx_rz_depth}
\end{figure}

For the second set of attacks, we fixed total rotation angles at $\theta=\pi$ and $\phi=\pi/2$. We vary the depth, or number of {\tt RX} and {\tt RZ} gates, while keeping the total equivalent rotation angles at a fixed sum of $\theta=\pi$ and $\phi=\pi/2$. For depth $d$, we use $d$ copies of {\tt RX($\pi$/d)} followed by $d$ copies of {\tt RZ($\pi$/2d)}. An example with $d=2$ is shown in Figure \ref{fig_attack_rx_rz_depth}.

We chose $\theta=\pi$ because, based on preliminary testing, it is the best non-zero rotation angle for the attacker. For $\theta=\pi$, $\phi=\pi/2$ is the choice of $\phi$ angle that is best for the attacker.

\subsection{Hiding Reset Operation Attack with {\tt CX} Gate Circuits}

Further, we considered circuits involving multiple qubits. We ran experiments with a series of {\tt CX} gates, using the victim qubit as the control qubit. {\tt CX} gates have long duration compared to single-qubit gates. While the control of the {\tt CX} gate does not affect the qubit state, allowing the attacker to gain information about the victim. The main goal is to evaluate the effect of time delay on the success of the attack. We hope to gain insight into whether duration of a circuit could be used to classify potentially malicious circuits.

\begin{figure}[t]
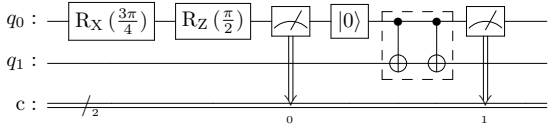

    \centering
    \include{figures/example_circuits/cx1}
    \caption{\small Example of circuit with {\tt CX} gates used as a masking circuit, different number of {\tt CX} gates can be tested for efficacy of the masking~circuit.}
    \label{fig_attack_cx}
\end{figure}

As shown in Figure~\ref{fig_attack_cx}, we repeat a number of {\tt CX} gates with the victim qubit, $q_0,$ as the control. The attacker only makes a measurement on the control qubit of the {\tt CX} gates.

\subsection{Hiding Reset Operation Attack with QASM Benchmarks} 

Aside from single-qubit masking circuits and circuits with {\tt CX} gates, an attacker may try more complex and deeper circuits to hide an attack. In particular, they could try to disguise their attack as a benign circuit, for example using some of the QASM benchmark circuits \cite{li2022qasmbench}. We evaluate whether it is possible for an attacker to perform a reset attack under our threat model using some common QASM benchmarks.

\subsubsection{2-Qubit Grover Search Circuit}

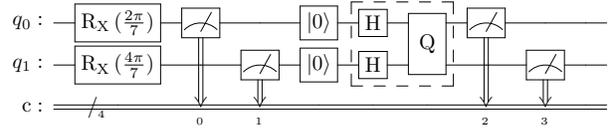
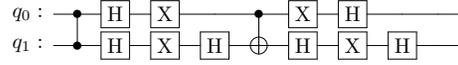
\begin{figure}[t]

\subfloat[2-qubit Grover circuit.]{\scalebox{\Scale}{
\centerline{\Qcircuit @C=\Width em @R=\RowHeight em @!R {
	 	\nghost{{q}_{0} :  } & \lstick{{q}_{0} :  } & \gate{\mathrm{R_X}\,(\mathrm{\frac{2\pi}{7}})} & \meter & \qw & \gate{\mathrm{\left|0\right\rangle}} & \gate{\mathrm{H}}  & \multigate{1}{\mathrm{Q}} & \meter & \qw & \qw & \qw\\
	 	\nghost{{q}_{1} :  } & \lstick{{q}_{1} :  } & \gate{\mathrm{R_X}\,(\mathrm{\frac{4\pi}{7}})} & \qw & \meter & \gate{\mathrm{\left|0\right\rangle}} & \gate{\mathrm{H}} & \ghost{\mathrm{Q}} & \qw & \meter & \qw & \qw \gategroup{1}{7}{2}{8}{.7em}{--} \\
	 	\nghost{\mathrm{{c} :  }} & \lstick{\mathrm{{c} :  }} & \lstick{/_{_{4}}} \cw & \dstick{_{_{\hspace{0.0em}0}}} \cw \ar @{<=} [-2,0] & \dstick{_{_{\hspace{0.0em}1}}} \cw \ar @{<=} [-1,0] & \cw & \cw & \cw & \dstick{_{_{\hspace{0.0em}2}}} \cw \ar @{<=} [-2,0] & \dstick{_{_{\hspace{0.0em}3}}} \cw \ar @{<=} [-1,0] & \cw & \cw }}}\label{2q_grover}}

\subfloat[2-qubit Grover circuit with operator Q~decomposed.]{\scalebox{\Scale}{\centerline{\Qcircuit @C=\Width em @R=\RowHeight em @!R { \\\nghost{{state}_{0} :  } & \lstick{{q}_{0} :  } & \ctrl{1} & \gate{\mathrm{H}} & \gate{\mathrm{X}} & \qw & \ctrl{1} & \gate{\mathrm{X}} & \gate{\mathrm{H}} & \qw & \qw & \qw\\\nghost{{state}_{1} :  } & \lstick{{q}_{1} :  } & \control\qw & \gate{\mathrm{H}} & \gate{\mathrm{X}} & \gate{\mathrm{H}} & \targ & \gate{\mathrm{H}} & \gate{\mathrm{X}} & \gate{\mathrm{H}} & \qw & \qw}}}\label{2q_grover_op}}

\caption{\small Example of using 2-qubit Grover circuit used as a masking circuit, circuits with different bitstrings and operators can be tested for efficacy of the masking circuit. The Hadamard, {\tt H}, gate can be realized using the basis gates discussed in the text.}
\label{fig_attack_2q_ga}
\end{figure}

We begin with the 2-qubit Grover search circuit. To start the search algorithm, the qubits need to be initialized into a uniform superposition with Hadamard gates. Then, the Grover operator, {\tt Q}, is applied to amplify the amplitude of the correct answer via rotations done by {\tt Q}. An example of 2-qubit Grover search is shown in Figure~\ref{2q_grover}.

We used Grover search with answer bitstring $11$. The circuit for the algorithm is boxed in Figure \ref{2q_grover}. The Grover operator {\tt Q} is decomposed in Figure \ref{2q_grover_op}. The attacker uses this circuit after the reset gates and before final measurement, like the previous attacks.

Unlike the single-qubit attack circuits, the attacker makes measurements on all involved qubits. The victim qubits are initialized with $\theta$ rotations independently of each other, that is, the rotation angles are not necessarily the same for each qubit. We limit the range of possible initial angles so that the total number of circuits for each trial does not exceed our limit on the \jakarta machine of $300$ circuits per job. For 2-qubit Grover, each qubit is initialized by the victim with a rotation of $\theta\in\{0,\frac{\pi}{7},\frac{2\pi}{7},\frac{3\pi}{7},\frac{4\pi}{7},\frac{5\pi}{7},\frac{6\pi}{7},\pi\}$.

\subsubsection{3-Qubit Grover Search Circuit}

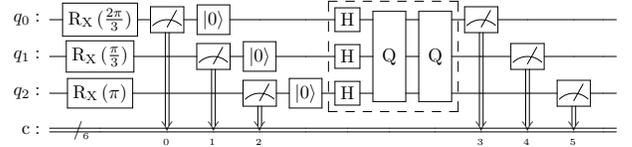
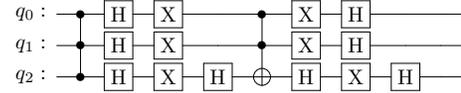
\begin{figure}[t]
\centering
\subfloat[3-qubit Grover circuit.]{
\scalebox{0.7}{
\centerline{\Qcircuit @C=0.7 em @R=\RowHeight em @!R {
	 	\nghost{{q}_{0} :  } & \lstick{{q}_{0} :  } & \gate{\mathrm{R_X}\,(\mathrm{\frac{2\pi}{3}})} & \meter & \gate{\mathrm{\left|0\right\rangle}} & \qw & \qw & \gate{\mathrm{H}} & \multigate{2}{\mathrm{Q}} & \multigate{2}{\mathrm{Q}} & \meter & \qw & \qw & \qw & \qw\\
	 	\nghost{{q}_{1} :  } & \lstick{{q}_{1} :  } & \gate{\mathrm{R_X}\,(\mathrm{\frac{\pi}{3}})} & \qw & \meter & \gate{\mathrm{\left|0\right\rangle}} & \qw & \gate{\mathrm{H}} & \ghost{\mathrm{Q}} & \ghost{\mathrm{Q}} & \qw & \meter & \qw & \qw & \qw\\
	 	\nghost{{q}_{2} :  } & \lstick{{q}_{2} :  } & \gate{\mathrm{R_X}\,(\mathrm{\pi})} & \qw & \qw & \meter & \gate{\mathrm{\left|0\right\rangle}} & \gate{\mathrm{H}} & \ghost{\mathrm{Q}} & \ghost{\mathrm{Q}} & \qw & \qw & \meter & \qw & \qw \gategroup{1}{8}{3}{10}{.7em}{--}\\
	 	\nghost{\mathrm{{c} :  }} & \lstick{\mathrm{{c} :  }} & \lstick{/_{_{6}}} \cw & \dstick{_{_{\hspace{0.0em}0}}} \cw \ar @{<=} [-3,0] & \dstick{_{_{\hspace{0.0em}1}}} \cw \ar @{<=} [-2,0] & \dstick{_{_{\hspace{0.0em}2}}} \cw \ar @{<=} [-1,0] & \cw & \cw & \cw & \cw & \dstick{_{_{\hspace{0.0em}3}}} \cw \ar @{<=} [-3,0] & \dstick{_{_{\hspace{0.0em}4}}} \cw \ar @{<=} [-2,0] & \dstick{_{_{\hspace{0.0em}5}}} \cw \ar @{<=} [-1,0] & \cw & \cw }}}\label{3q_grover}}

\subfloat[3-qubit Grover circuit with operator Q~decomposed.]
{
\scalebox{\Scale}{\centerline{\Qcircuit @C=\Width em @R=\RowHeight em @!R { \nghost{{q}_{0} :  } & \lstick{{q}_{0} :  } & \ctrl{1} & \gate{\mathrm{H}} & \gate{\mathrm{X}} & \qw & \ctrl{1} & \gate{\mathrm{X}} & \gate{\mathrm{H}} & \qw & \qw & \qw\\ \nghost{{q}_{1} :  } & \lstick{{q}_{1} :  } & \ctrl{1} & \gate{\mathrm{H}} & \gate{\mathrm{X}} & \qw & \ctrl{1} & \gate{\mathrm{X}} & \gate{\mathrm{H}} & \qw & \qw & \qw\\ \nghost{{q}_{2} :  } & \lstick{{q}_{2} :  } & \control\qw & \gate{\mathrm{H}} & \gate{\mathrm{X}} & \gate{\mathrm{H}} & \targ & \gate{\mathrm{H}} & \gate{\mathrm{X}} & \gate{\mathrm{H}} & \qw & \qw\\ }}}
\label{3q_grover_op}
}

\caption{\small Example of using 3-qubit Grover circuit used as a masking circuit, circuits with different bitstrings and operators can be tested for efficacy of the masking circuit. The Hadamard, {\tt H}, gate can be realized using the basis gates discussed in the text.}
\label{fig_attack_3q_ga}
\end{figure}

We also experimented with the 3-qubit Grover search circuit, which looks similar to 2-qubit Grover search, but has more gates and is deeper. Each qubit is initialized by the victim with a rotation of $\theta\in\{0,\frac{\pi}{3},\frac{2\pi}{3},\pi\}$. An example of 3-qubit Grover search is shown in Figure~\ref{fig_attack_3q_ga}.

\subsubsection{Random Number Generator Circuit}

From the QASM Benchmark suite, there are two small-scale circuits that do not use multi-qubit gates, namely, the quantum random number generator, and the inverse Quantum Fourier Transform (QFT). However, the inverse QFT circuit requires conditional operations, which are currently unavailable on IBM Quantum machines. So we consider the random number generator on $4$~qubits.

\begin{figure}[t]
    \centering
    \input{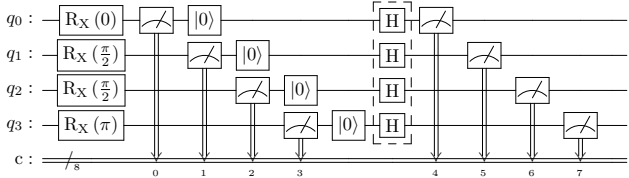}
    \caption{\small Example of Quantum Random Number Generator (QRNG) used as a masking circuit. The Hadamard, {\tt H}, gate can be realized using the basis gates discussed in the text.}
    \label{fig_attack_qrng}
\end{figure}

The Quantum Random Number generator, shown in Figure~\ref{fig_attack_qrng}, uses Hadamard gates to produce a uniform superposition before measurement. This attacker circuit has the smallest depth of the benchmarks tested by this paper, with a depth~$1$.

\section{Reset Operation Error Channel Analysis}
\label{sec_reset_error}

\begin{figure*}[t]
\centering

\subfloat[\small Evaluation on real \jakarta machine.]{%
  \includegraphics[trim={0 0 0 1.1cm},clip,width=0.85\textwidth]{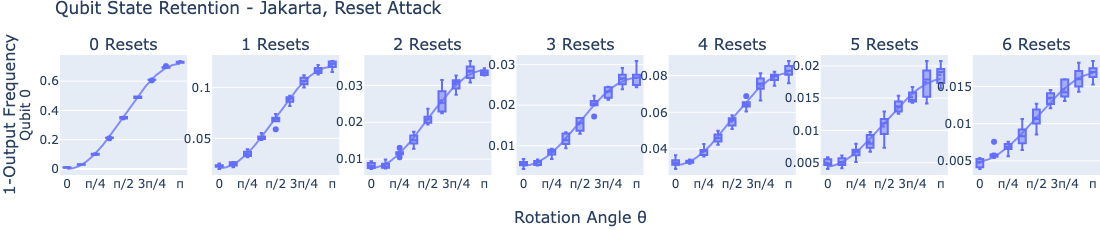}%
  \label{real_reset}
}
\linebreak
\subfloat[\small Simulator evaluation, using built-in simulated reset operation.]{%
  \includegraphics[trim={0 0 0 1.1cm},clip,width=0.85\textwidth]{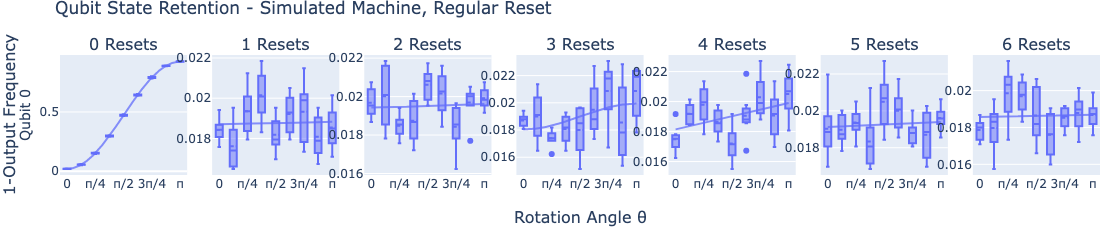}%
  \label{sim_reset}
}
\linebreak
\subfloat[\small Simulator evaluation, using ``measurement + {\tt X} gate'' approach to emulate reset operation.]{%
  \includegraphics[trim={0 0 0 1.1cm},clip,width=0.85\textwidth]{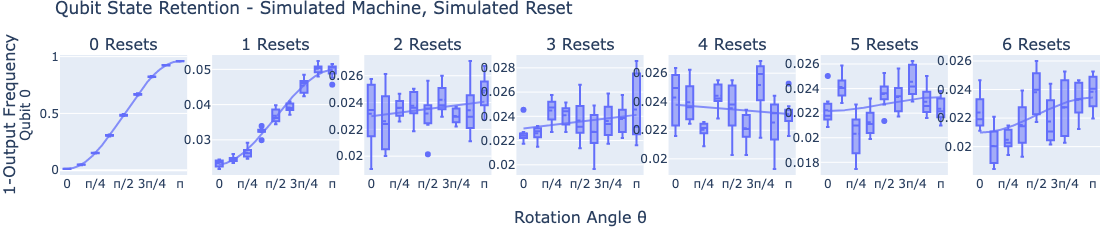}%
  \label{reset_m_x}
}

\caption{\small Qubit state retention, comparison of: (a) reset operation on real machine, (b) simulated reset operation, and (c) simulated reset operation using ``measurement + {\tt X} gate'' approach.}
\label{compare_sim}
\end{figure*}

Before we present evaluation of the different attacks that use masking circuits, we discuss characteristics of the reset operation. Further, we compare behavior of the reset operation on real \jakarta machine to two types of simulation to motivate our use of real \jakarta for subsequent~evaluation.

\subsection{Behavior of Reset Operation}
\label{sub_sec_sigmoid}
Qubits are often implemented with $\ket{1}$ as a higher energy state than $\ket{0}$. This results in a higher probability of an incorrect readout for qubit in state $\ket{1}$ compared to state $\ket{0}$. Thus, we expect states with a higher amplitude of $\ket{1}$ to have a higher probability of being the $\ket{1}$ state after a reset \cite{mi2022securing}. This error of real machine resets is seen in Figure \ref{real_reset}, and allows the attacker to extract information about the $\theta$ angle of the victim qubit based on the measured $1$-output frequency~\cite{mi2022securing}.

Given the state: $$\ket{\psi}=\cos\left(\frac{\theta}{2}\right)\ket{0}+e^{i\phi}\sin\left(\frac{\theta}{2}\right)\ket{1},$$ recall that the probability of measuring $1$ is $\sin^2\left(\frac{\theta}{2}\right)$ according to the Born rule interpretation. This motivates an error channel characterization \cite{mi2022securing} based on the probability of measuring $1$ post-reset: $$E(\theta)=a\left(b\sin^2\left(\frac{\theta}{2}\right)+(b-1)\frac{\theta}{\pi}\right)+c,$$ where $a\in[-1,1],b,c\in[0,1]$. On the domain, $\theta\in[0,\theta],$ the output probability looks like a sigmoid curve. This is seen in Figure~\ref{real_reset}.
This error channel parameterization is important to our attack evaluation in Section \ref{sec_results}.

\subsection{Observed Fidelity Improvements of Reset Operations}

Over the past year, IBM machines have improved in fidelity and yield lower error rates. Indeed, according to IBM's reported error rates through Qiskit's \texttt{IBMQBackend.properties()} method, we found that for qubit $0$ of \jakarta, readout error rate has dropped from $0.0360$ to $0.0218$ over the past year. In addition, the rate of measuring $0$ from a $\ket{1}$ state dropped from $0.0464$ to $0.0340$, and the rate of measuring $1$ from a prepared $\ket{0}$ state dropped from $0.0256$ to $0.0096.$

While the error rates and noise have decreased, the current experimental results suggest that the same reset error based on amplitude of $\ket{1}$ is still present in IBM machines. In comparison to last year, the $1$-output frequency of an attacker measuring the victim qubit after $6$ resets still displays a significantly higher frequency for $\theta=\pi$ than for $\theta=0.$ At the same time, the noise is of much smaller magnitude, as indicated by the smaller error bars.

With decreasing noise to signal ratio, the possibility of a reset error channel attack is becoming actually greater. The attacker is able to recover more information from the victim with ever-increasing probability, even after numerous reset~operations.

\subsection{Study of Simulated vs. Real Reset Operations}
\label{sub_sec_sim_reset}

We compared different types of simulated reset operations with the real \jakarta machine. We used \texttt{AerSimulator}, with a noise model directly imported from IBM’s \jakarta backend. In theory, the simulator should behave as the real backend for all qubit gates. Based on our testing, the built-in simulated reset operation does not have the same error as the real machine’s reset operation. While the real reset operation has a higher probability of an incorrect reset for qubits with a larger magnitude of $\ket{1}$, the simulated reset removes this: there is no clear correlation between the victim qubit’s original theta angle and the output frequencies post-reset. The data is shown in Figure \ref{sim_reset}.

Given the built-in simulated reset operation does not behave as a real one, we then attempted replacing the built-in reset operation with a measurement followed by an {\tt X} gate conditioned on the measurement being $1$ -- this should in theory represent the behavior of the reset operation. We did observe more realistic results in the case of $1$ reset, as the sigmoid shape can be seen in Figure \ref{reset_m_x}. However, the addition of two or more reset operations with the simulator results in noisy data, and no longer fits a sigmoid curve. This suggests that the simulated reset does not emulate the real machine when using a measurement followed by an {\tt X} gate as the reset operation.

Both the simulator's built-in simulated reset operation and the measurement followed by {\tt X} gate scheme on the simulator produce a lot of noise: the $1$-output frequencies vary a lot depending on the victim qubit’s $\phi$ angle compared to the real machine. At this time, the simulator is unable to accurately replicate the behavior of the reset operation on IBM Quantum machines, and our evaluation in the rest of the paper users' data from real \jakarta machine.

\section{Evaluation of Efficacy of Masking Circuits }
\label{sec_results}

\begin{figure*}[h]
\centering
  \includegraphics[width=0.85\textwidth]{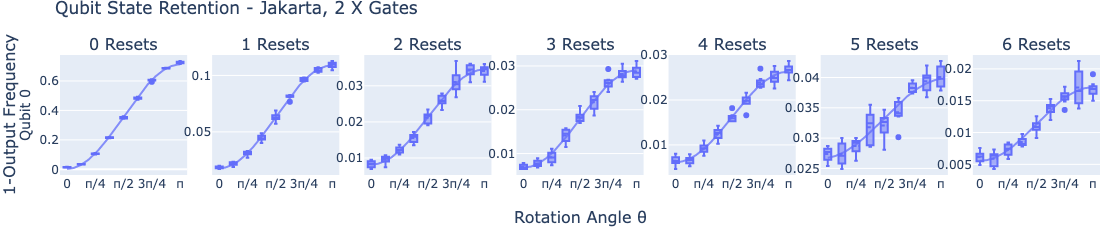}
  \caption{\small Example $1$-output frequency of {\tt X} gate masking circuit. Circuits with $0,2,4,6,8,16,$ and $32$ {\tt X} gates were used as the attacker circuit. Experiments done with qubit $0$ of \jakarta. Only results for $2$ {\tt X} gates are shown, with the other graphs having a similar~shape.}
  \label{6_resets}
\end{figure*}

In this section, we present evaluation results for different masking circuits previously discussed in Section~\ref{sec_attack_model}. The masking circuit evaluation is based on: 1) {\tt X} gates, 2) {\tt RX} and {\tt RZ} gates, 3) {\tt CX} gates, and 4) QASM benchmarks. For all circuits, we ran experiments on \jakarta using a varying number of reset gates after the victim and a varying circuit depth for the masking circuits, where possible.

\subsection{Evaluation Metrics}

To evaluate the effectiveness of each attack circuit, we use a metric of signal-to-noise ratio (SNR). We computed the SNR to estimate how much information the attacker could extract from the output frequency data when different types of masking circuits are~used.

We compute the error channel characterization parameter $a$, which represents the amplitude of our sigmoid fit. The fit is described in Section~\ref{sub_sec_sigmoid}. We compute the standard deviation in 1-output frequency for each fixed $\theta$ as $\phi$ varies. Finally, we compute the average standard deviations over all input $\theta$ values, denoted $\sigma$. Then the signal-to-noise ratio is defined as $a/\sigma$, expressed on a log scale (decibels).

\subsection{Reset Schemes}

\begin{figure}[t]
\centering
\subfloat[\small Evaluation on real \jakarta machine.]{%
  \includegraphics[width=\columnwidth]{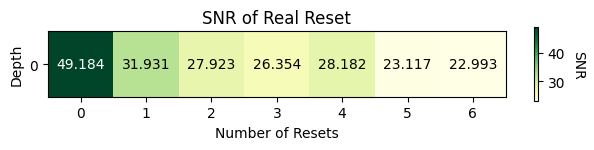}%
  \label{snr_real_reset}
}
\linebreak
\subfloat[\small Simulator evaluation, using built-in simulated reset operation.]{%
  \includegraphics[width=\columnwidth]{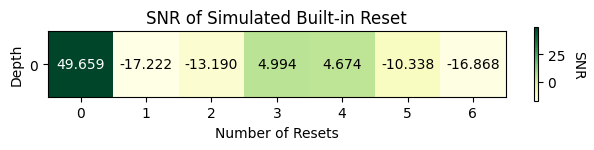}%
  \label{snr_sim_reset}
}
\linebreak
\subfloat[\small Simulator evaluation, using ``measurement + {\tt X} gate'' approach to emulate reset operation.]{%
  \includegraphics[width=\columnwidth]{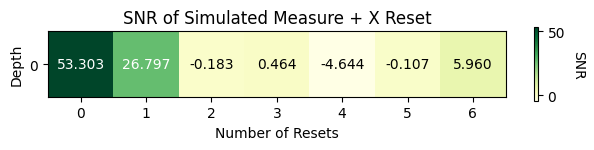}%
  \label{snr_reset_m_x}
}

\caption{\small Comparison of SNR for reset on real machine, simulated reset, and simulated reset using ``measurement + {\tt X} gate'' approach.}
\label{compare_sim_SNR}
\end{figure}

Using this metric, we can compare the different reset schemes described in Section~\ref{sub_sec_sim_reset}.
Figure~\ref{compare_sim_SNR} shows the SNR for the three different reset schemes. The SNR metric aligns with the analysis of Section~\ref{sub_sec_sim_reset}. We observe a relatively strong SNR for the real reset. For the simulated reset, there is a sharp decline in SNR after adding the first reset. Using a measurement and X gate to simulate reset, the SNR for one reset is relatively high, but adding more resets decreases the SNR drastically.

\subsection{Attack Involving Identity Circuits}

\begin{figure}[h]
\includegraphics[width=0.94\columnwidth]{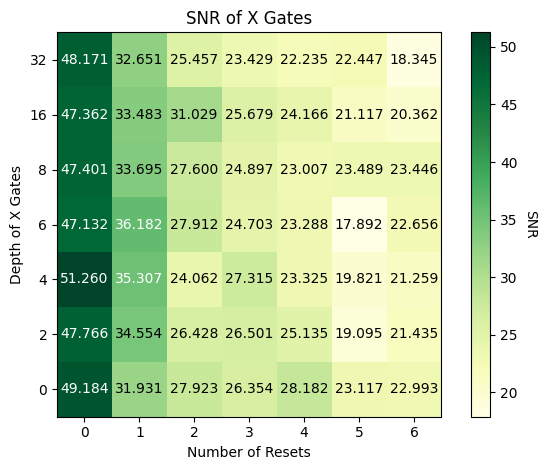}
\caption{\small SNR for {\tt X} gate masking circuit experiments. A series of up to $32$ {\tt X} gates were tested.}
\label{4x_snr}
\end{figure}

We ran circuits with a series $0,2,4,6,8,16,$ and $32$ X gates as the attacker circuit. For each attack circuit, we added up to $6$ reset gates after the victim. All experiments were run on qubit $0$ of \jakarta.

Figure \ref{6_resets} displays the 1-output frequency of each attack circuit as a function of the victim qubit's rotation angle $\theta.$ For the purposes of conserving space, only the results for $2$ X gates are shown. The graphs for more resets display the same sigmoid shape.

We expect that as the depth of the circuit increases or the number of reset gates, the attacker's job becomes harder as more noise is introduced. Figure \ref{4x_snr} shows the SNR plotted on a decibel scale for all depths of $X$ gate circuits and all numbers of reset gates. As expected, increasing the number of resets results in decreasing the signal-to-noise ratio. The correlation coefficient between these two variables is $-0.862$, indicating a strong negative correlation. The most significant decrease in SNR resulted from the addition of the first reset gate, with subsequent resets having a lesser effect on SNR.

The depth of the circuit, measured as number of $X$ gates, did not appear to have much effect on the SNR, as there is no clear trend of the SNR as depth increases. The correlation coefficient between these two variables is $-0.057$, indicating no significant correlation.

\subsection{Attack Involving {\tt RX} and {\tt RZ} Gate Circuits}
 
\begin{figure}[ht!]
\centerline{\includegraphics[width=\columnwidth]{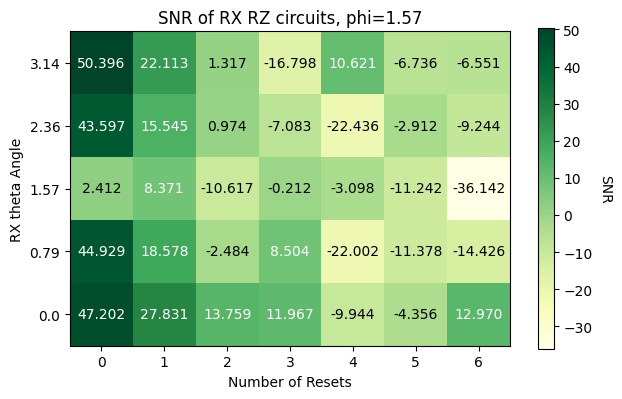}}
\caption{\small SNR for the first set of {\tt RX} and {\tt RZ} attacker experiments. The rotation angles were varied while the depth was fixed at $1$ of each gate.}
\label{rx_angle_snr}
\end{figure}

In the first set of experiments, we used $\theta\in\{0,\frac{\pi}{4},\frac{\pi}{2},\frac{3\pi}{4},\pi\}$ and $\phi \in \{0, \frac{\pi}{2}, \pi, \frac{3\pi}{2}\}$ for the attacker's {\tt RX} and {\tt RZ} gates, respectively. We observed that $\phi=\pi/2$ seems particularly beneficial for the attacker compared to other $\phi$ angles. The results for this $\phi$ angle are shown in Figure \ref{rx_angle_snr}.

For $\theta = \pi/2$, the SNR is the lowest, meaning it is the most difficult for the attacker to extract information about the victim's initial angle. This coincides with our expectation, because after an {\tt RX} rotation by $\pi/2$, both initial states $\ket{0}$ and $\ket{1}$ have the same output probability of $\frac{1}{2}.$

As the $\theta$ angle changes from $\pi/2$ towards $0$ or $\pi$, it becomes easier for the attacker to distinguish the victim's initial state. Increasing the number of resets generally decreases the signal-to-noise ratio, as expected.

\begin{figure}[ht]
\centerline{\includegraphics[width=\columnwidth]{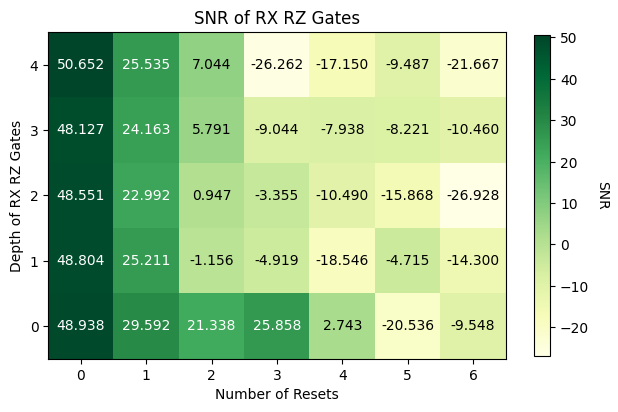}}
\caption{\small SNR for the first set of {\tt RX} and {\tt RZ} attacker experiments. The rotation angles were varied while the depth was fixed at $1$ of each gate.}
\label{rx_depth_snr}
\end{figure}

We then experimented by varying the depth of {\tt RX} and {\tt RZ} gates, while keeping the total rotation angles at $\theta = \pi$ and $\phi = \pi/2.$ Each rotation gate used the same $\theta$ or $\phi$ angle. For example, for depth $2$ we used two {\tt RX($\pi/2)$} gates and two {\tt RZ($\pi/4)$} gates. We ran a control group with no attacker, labelled depth $0$ in Figure \ref{rx_depth_snr}.

For $3$ resets, increasing the depth decreases the SNR. However, for $2$ resets, the opposite effect occurs. In general, the correlation between depth and SNR is $-0.14$, indicating little to no correlation.

\subsection{Attack Involving {\tt CX} Gate Circuits}
We experimented with a series of CX gates as the attacker. We used qubit $0$ on \jakarta as the victim qubit, and we added up to $6$ CX gates in series after the reset gates, using the victim qubit, qubit $0$, as the control qubit.

\begin{figure}[t]
\includegraphics[width=\columnwidth]{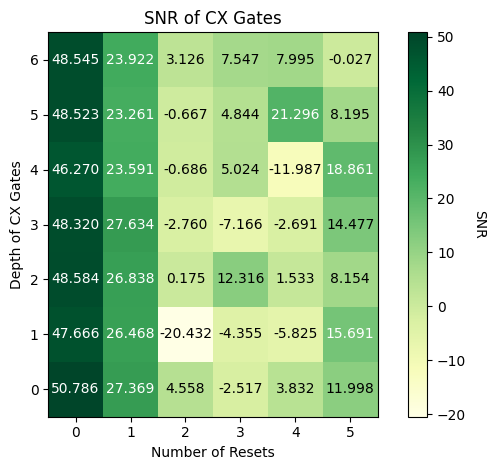}
\caption{\small SNR for {\tt CX} gate attacker experiments. The {\tt CX} gates were used qubit $0$ as the control qubit. Output results and SNR are based on qubit $0$.}
\label{cx_snr}
\end{figure}

Interestingly, increasing the number of reset gates from $0$ to $1$ or from $1$ to $2$ decreases the SNR, while the increasing the number of reset gates beyond $2$ seems to increase the SNR, on average. 
For any number of reset gates, the depth of the {\tt CX} gates does not have strong correlation with the SNR, with a correlation coefficient of $0.039.$

Due to numerous job requests, the circuits for this set of experiments were executed over several days. This may have introduced noise in the data, as IBM Q machines have slightly different error rates across different execution times.

\subsection{Attack Involving Grover search Circuits}

To compute the signal-to-noise ratio with a multi-qubit circuit, we need a new measure of signal. For each qubit, we consider the $1$-output frequency as a function of all qubits' initial angles. We compute the sum of the squares of the gradients with respect to each input dimension, then take a square root. This final value, the Root-Mean-Square (RMS) gradient, is roughly a measure of the rate of change in $1$-output frequency as we change the input angles. As a measure of noise, we use the average standard deviation in output frequency, as in the single-qubit case. For each combination of initial angles, we did $8$ trials. We compute the quotient as the SNR for each qubit. 

\begin{figure}[t]
\includegraphics[width=\columnwidth]{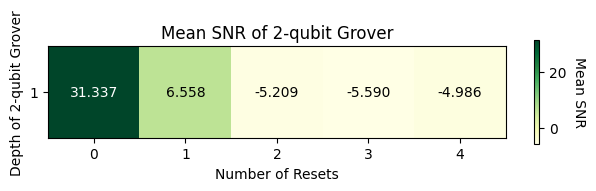}
\caption{\small SNR for 2-qubit Grover circuit experiments. Average gradient is used as the measure of signal for calculating the SNR.}
\label{2_ga_snr}
\end{figure}

Figure \ref{2_ga_snr} shows the results for $2$-qubit Grover search.
We observed sharp declines in SNR after $1$ and $2$ resets. Increasing the number of resets past $2$ does not appear to significantly impact the SNR.

\begin{figure}[t]
\includegraphics[width=\columnwidth]{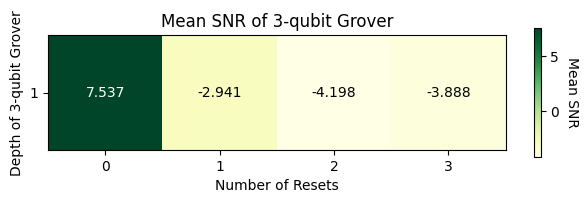}
\caption{\small SNR for 3-qubit Grover circuit experiments. Average gradient is used as the measure of signal for calculating the~SNR.}
\label{3_ga_snr}
\end{figure}

Figure \ref{3_ga_snr} shows the results for $3$-qubit Grover search. 

We observed a sharp decline in SNR after $1$ reset. Increasing the number of resets past $1$ does not appear to significantly impact the SNR. We also note the difficulty of drawing a conclusion given the limited data we have, especially for $3$-qubit Grover's.

\subsection{Attack Involving QASM Benchmark Circuits}

Below are results for the QRNG circuit on four qubits. We used an initial rotation angle of $\theta\in\{0,\pi/2, \pi\}$ for each qubit. For every combination of initial angles, we ran $6$ trials.

\begin{figure}[t]
\includegraphics[width=\columnwidth]{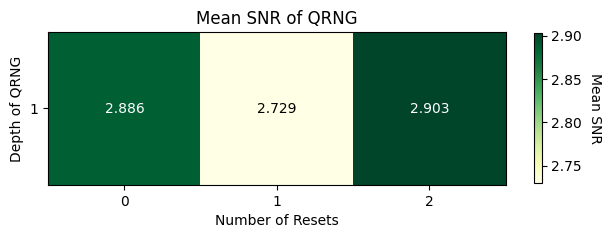}
\caption{\small SNR for QRNG benchmark circuit experiments. Hadamard gates on each qubit are used to achieve a uniform, random output. Average gradient is used as the measure of signal for calculating the~SNR.}
\label{qrng_snr}
\end{figure}

For three or more resets, the IBM computers ran into internal error. This error also appeared for \jakarta for large numbers of resets on the Grover search algorithms. 

Figure \ref{qrng_snr} represents the mean SNR of all four qubits of the QRNG circuit. Interestingly, increasing the number of reset gates up to $2$ does not seem to have a significant impact on the SNR. 

\subsection{Summary of the Attacks and the Evaluation}

We have shown that for single-qubit gates used in the masking $C$ circuit, the attacker may use simple identity circuits consisting of pairs of {\tt X} gates, or circuits consisting of  {\tt RX} and {\tt RZ} gates. For multi-qubit gates, an attacker can also try to hide the attack by using masking circuit with {\tt CX} gates, as long as the target qubit is the control qubit of the {\tt CX} gate. We also showed conditions under which the attack becomes more difficult, such as when qubits are targets of {\tt CX} gate. We confirm our expectation by running select QASM benchmark circuits, and showing that it is difficult for the attacker to leak the victim's state, due to the presence of multi-qubit gates or other non-identity gates, if the masking circuit $C$ is a full QASM benchmark, for example. Based on these findings, a defense for our extended reset operation attack can be developed.

\section{Defense Against the New Reset Operation Attacks}
\label{sec_defense}

We provide a number of compile-time heuristics that can be used to detect possibly malicious attacks that try to use masking circuits with a measurement to perform a reset operation attack. Our compile-time solution is complimentary to the existing ``secure reset'' work~\cite{mi2022securing}, which is a run-time solution. Further, our approach is different from the existing quantum computer antivirus~\cite{deshpande2023design}, which focuses on the exact quantum circuit pattern matching.

\subsection{Detecting Attacks that use Identity Circuits}
\label{sec_defense_identity}

In the case that the attacker places an identity circuit before the measurement, we scan all gate operations done after the last reset gate and before the final measurement. We use Qiskit's {\tt Operator} class to convert any potential adversarial circuit into its matrix representation. Then, we check if this matrix is an identity. This is efficient for circuits with a small number of qubits. For large circuits, we can loop through each qubit and check the gates that operate on it. If these are single-qubit gates only, and if these operations are equivalent to identity, our program flags the circuit as suspicious.

If a circuit consists of an identity followed by measurement, our program will flag it as suspicious. The size of the matrix  representation scales exponentially with the number of qubits involved, so it is limited to smaller circuits. In testing, we generated $100$ random $7$-qubit circuits of depth $10$, and our program successfully and efficiently flagged all of these as identity circuits.

\subsection{Heuristics for General Attack Detection}

In the most general case, the attacker may use a non-identity circuit as a masking circuit, or he or she may use many qubits that make matrix representations infeasible to work with. In this case, we present an approach that considers each qubit one at a time.

For each qubit, we can compute the matrix representation of all gates involving the specific qubit. We first check if the qubit is involved in any multi-qubit gates. Based on our results, circuits involving multi-qubit gates are not susceptible to the reset gate attacks. However, single-qubit gates introduce little error, and even at large depths, the attack can still extract information on these qubits. Thus, any qubits involved in only single-qubit gates, or the control qubit of a {\tt CX} gate, will be noted by our program.

In the case that a qubit is only involved in single-qubit gates, our program checks if the circuit applies an effective {\tt RX} rotation on the qubit. Based on our results, an effective {\tt RX} rotation close to $\pi/2$ makes it difficult for the attacker to perform the attack. So, we propose flagging any qubit with effective rotation $\theta>3\pi/4$ or $\theta<\pi/4.$

Note that for most circuits, most qubits will have more complex operations that cannot be reduced to an equivalent {\tt RX} rotation. In this case, our program can still note whether the qubit is effectively identity, or only involves single-qubit~gates.

\subsection{Implementation}

We assume our program has access to the circuit that is to be checked, e.g., our program can be used by IBM to scan submitted circuits before they execute on the quantum computers. Given an input circuit, it is simple to count the circuit depth of the possibly malicious input circuit. Additionally, Qiskit provides functionality to convert circuits into their matrix representation. Since the number of resets used is controlled by the quantum computer provider, we assume the number of resets is an input or configuration given to our~program.

To scan circuits, we first extract the gates from the input quantum circuit, and for any given qubit, check if the gate operates on the qubit. If so, we save the instruction for the gate. In the end, we make a quantum circuit from the list of instructions, yielding the subset of the original circuit that involves each specific qubit. On this smaller circuit, we compute the matrix representation and check for existence of multi-qubit gates, equivalence to identity, and equivalence to a single {\tt RX} rotation.

Based on our testing, for attacker circuits of $32$ {\tt X} gates, $6$ {\tt CX} gates, $2$-qubit Grover, $3$-qubit Grover, and the QRNG Benchmark, our antivirus program can complete a scan in $0.017$ seconds, $0.009$ seconds, $0.024$ seconds, $0.130$ seconds, and $0.017$ seconds, respectively.

\section{Related Work}
\label{sec_related_workd}

Considering attacks on quantum computers, the closest related work is the work which analyzed attacks on reset operations~\cite{mi2022securing}. The authors showed for the first time that imperfections in reset operations can lead to possible information leaks between shots of circuits. Our work extends this prior work and shows more advanced attacks where use of masking circuit is used to help hide the attacker while still allowing for information leak to be extracted by the attacker.

Considering protections for quantum computers, previous work has suggested an ``antivirus'' programs which can be used to detect malicious quantum circuits. The authors used a directed acyclic graph (DAG) to represent an input quantum circuit. In the DAG with non-commutativity (DAGNC) representation, a quantum circuit can be represented as a multigraph. Vertices in the multigraph correspond to gates in the quantum circuit, and edges correspond to orders between gates. The edge from node $i$ to $j$ means that the gates corresponding to node $i$ and $j$ have at least one qubit or classical bit in common, and the gate corresponding to node $i$ executes before the gate corresponding to node $j$. The authors used this representation to find instances of smaller ``virus'' circuits in the larger input quantum circuit. In contrast, our work does not require use of DAG, but instead scans individual qubits and computes the matrix form of the input circuit. Our defense program could be incorporated into the antivirus as a new feature.

\section{Conclusion}
\label{sec_conclusion}

In this work, we demonstrated how a set of new, extended reset operation attacks could lead to critical information leakage from quantum programs executed on quantum computing cloud environments. This work showed that this new kind of reset operation attack could be more stealthy than the previous reset operation attacks, by hiding the intention of the attacker's circuit. The work evaluated how an attacker can mask the circuit by adding simple identity circuits or non-identity circuits consisting of {\tt RX} and {\tt RZ} gates for single-qubit gates or {\tt CX} gates. This work also showed that more complex circuits may render the attack difficult. Based on the findings, this work showed a set of heuristic defenses that could be applied at compile time to check and flag the new kind of malicious circuits.

\balance

\bibliographystyle{IEEEtranS}
\bibliography{main}

\end{document}